\documentclass[traditabstract]{aa} 
\usepackage{graphicx}
\usepackage{amssymb, amsmath}
\usepackage{amsmath}
\usepackage{graphicx}
\usepackage{ulem}
\usepackage{amsmath}
\usepackage{graphicx}
\usepackage{url}

\usepackage[usenames]{color}

\newcommand{\beq}{\begin{equation}}
\newcommand{\eeq}{\end{equation}}
\newcommand{\bea}{\begin{eqnarray}}
\newcommand{\eea}{\end{eqnarray}}

\newcommand{\ass}[1]{{Ap.\ Space Sci.,} {\bf #1}}

\renewcommand{\sp}[1]{{Solar Phys.} {\bf #1}}

\begin{document}

\title{The formation of (very) slowly rotating stars}
\titlerunning{Slowly rotating stars}

\author{H.C.\ Spruit}
\authorrunning{H.C.\ Spruit}
\institute{Max-Planck-Institut f\"{u}r Astrophysik,
  Karl-Schwarzschild-Str.\ 1,
  D-85748 Garching, Germany}
\date{\today}

\abstract{The slow rotation of some young stars and the extreme rotation periods of some Ap stars have so far defied explanation. The absence of sufficiently efficient braking mechanisms for newly formed stars points to the star formation process itself as the origin. I find that a mode of star formation exists by which a protostar can form  without accreting angular momentum. It depends on the survival of a magnetic connection between the accreting matter and the birth cloud. The conditions for this process to operate are analyzed, and illustrated with a generic (scale-free) model. Depending on the initial rotation rate of the accreting matter, either a Keplerian disk forms, or the gas ends up rotating with the rotation period of the cloud, even if it is at a large distance. The boundary in parameter space between these two outcomes is  sharp. 
 
}

\maketitle

\section{Introduction}

Newly formed main sequence stars show a remarkable range of rotation rates. Solar type stars in young clusters of age 3-5 My, for example, have periods ranging from half a day to 15d or more (Littlefair+ 2010), a factor 30.  In the middle-aged Pleiades cluster (125 My, Stauffer+ 1998), rotation periods are still in this range (Rebull+ 2016). Accommodating such a wide range presents a problem for explanations of stellar rotation rates. 

A second mystery is the long period end of the observed distribution. In the standard interpretation (`gyrochronology') a star with a rotation period of 15d would be a Gigayear old, rather than 3-5 My. The distribution is even more extreme in the case of A stars, in particular the magnetic Ap stars.  Measured rotation periods of these stars vary from half a day to decades. The record holder is $\gamma$ Equ ($P=97$ yrs: Bychkov+ 2016), while Mathys (2017) infers the likely existence of periods of centuries or more. Such long rotation periods  require explanations beyond current ideas.  As elsewhere, the extremes of a distribution can be more telling about its origin than its average or median. From this perspective, existing models  are discussed in sections \ref{lock} and \ref{wind} below. Sect.\ \ref{loss} then explains the possibility of protostars accreting mass  with negligible angular momentum, with a model developed in sects.\  \ref{msa} and \ref{magacc}.

\section{Braking mechanisms}
\label{brake}
After forming from an accretion disk, a pre-main sequence star is expected to rotate rapidly. Sun-like stars  would have a strong dynamo-generated field in their convective envelope, with a surface field strength $B_0$ of the order of a kG, while Ap stars a fossil field with a median dipole strength of order 3\,kG. Near the star the surface field continues outside as a potential field and corotating with it. The  strength $B_{\rm p}$ of its dipole component varies with distance $r$ as $B_{\rm p}(r)\sim B_0 (R/r)^3$, where $R$ is the star's radius and $B_0$ the surface value of $B_{\rm p}$.  Corotation of the field ends at some distance $r_{\rm A}$, either at the point where the field becomes affected by an external medium, or at the Alfv\'en surface of a magnetically powered wind of the star. In both cases, an azimuthal field component  $B_\phi(r)$ transmits a braking torque to the star.   $B_\phi(r)$  is small compared with $B_{\rm p}$ close to the star, but  it decreases with distance more slowly, and  eventually becomes comparable to  $B_{\rm p}$. This happens at  the Alfv\'en distance  $r_{\rm A}$, the point where the Alfv\'en speed matches the rotation speed of the field line. The torque $\cal T$ acting there  is transmitted to the star as a braking torque.  As in the case of a magnetic stellar wind or a pulsar magnetosphere, it is of the order  
\beq {\cal T}\approx r_{\rm A}^3 B_{\rm p,A}^2(r_{\rm A}).\eeq If $\tau_{\rm s}$ is the characteristic time scale $\Omega/\dot\Omega$ on which a star rotating at a rate $\Omega$ spins down under a torque $\cal T$: 
\beq I\, \Omega/\tau_{\rm s}= {\cal T}. \eeq
With the moment of inertia $I=k\,M R^2$, where $M$ and $R$ are the star's mass and radius and $k$ its dimensionless gyration constant, this can be expressed in terms of an estimate for the distance $r_{\rm A}$ where the torque must be acting in order to yield this time scale:
\beq 
({r_{\rm A}\over R})^3={3\over k}{\omega_{\rm A}^2\tau_{\rm s}\over \Omega},\label{rmR}
\eeq
where $\omega_{\rm A}$ is the  star's internal Alfv\'en frequency $\omega_{\rm A}=\bar v_{\rm A}/R$, with  $\bar v_{\rm A}=B_0/(4\pi\bar\rho)$, and $\bar\rho$ the mean density of the star. As an example assume a star with mean density $\bar\rho\approx 1$\,g/cm$^3$, a radius $R=10^{11}$ cm, field strength $B_0=1$ kG and $k=0.1$, rotating with a period of one day, and assume we want it to spin down on an (initial) time scale of $10^6$\,yr. Eq.\ (\ref{rmR}) then yields $r_{\rm A}/R\approx 5$\,: the braking agent has to be close to the star in this case. If instead we are interested in the last bit of rotation of an Ap star of $3\,10^3$\,G rotating at a period of 10\,yr, and require this rotation to disappear also on a PMS time scale of $10^6$\,yr, we get $r_{\rm A}\approx 1$ AU. In other words it would require the unlikely presence at this distance of an essentially non-rotating braking agent, acting over a period of  $10^6$\,yr. To derive more from these estimates, we need to consider the nature of the external medium. This is done in the next subsections.

\subsection{disk locking}
\label{lock}
In the idea of `disk locking' (for a critical review see Davies 2015) the external medium providing a braking torque is the disk through which the star is accreting its mass. The star's magnetic field defines a boundary, the magnetosphere, outside of which accretion takes place as a disk approximately unaffected by the field, with its inner edge at the radius $r_{\rm m}$ of the magnetosphere. Inside $r_{\rm m}$ the accretion takes place in a magnetically dominated form of Rayleigh-Taylor interchange (Spruit \& Taam 1990, Stehle+ 2001, Igumenshchev+ 2003, Marshall+ 2018). For the torque acting on the star to be negative (spindown), the orbital rotation rate of the inner edge, $\approx \Omega_{\rm K}(r_{\rm m})$, has to be less than the star's rotation $\Omega_*$. 
Combining this with (\ref{rmR}), this yields
\beq {r_{\rm m}\over R}<({3\,\omega_{\rm A}^2\tau \over k\Omega_{\rm K*}})^{2/3}, \eeq
where $\Omega_{\rm K*}=(GM/R^3)^{1/2}$ is the Kepler frequency at the surface of the star. For a $1M_\odot$ PMS star with $R=10^{11}$\,cm and a dipole field of 1\,kG, this yields $r_{\rm m}/R < 20$. The corresponding rotation period, the longest for which disk locking could work, is $\approx 20$\,d. For an Ap star of mass $2\,M_\odot$, radius $2\,R_\odot$ and dipole field strength $3000$\,G, the maximum rotation period is similar. 

For solar-type stars, however, a (dynamo-generated) field strength of 1 kG would be realistic only for rapidly rotating stars. The magnetic activity of stars with surface convection zones decreases with rotation period. The dipole component of the Sun ($P=25$ d) for example is only about 10\,G (Cameron+ 2010). This is too weak for significant interaction with an accretion disk, at any distance from the star. The long rotation periods seen in young clusters require a different explanation. The situation is better for the Ap stars since their fossil fields decay only on evolutionary time scales (cf. Fossati+ 2016). In this case disk locking could in principle work for periods up to a month, but periods as long as years or more again need another explanation. 

\subsection{magnetic winds}
\label{wind}
\subsubsection{disk winds}
The presence of slowly rotating stars in young clusters may be interpreted as the result of a strong angular momentum loss process during their formation. The two options proposed in the literature are loss trough a magnetically driven stellar wind (St{\c e}pie{\'n} 2000) or loss by a disk wind (e.g.\ Davies 2015). A well-studied model is a disk wind powered by an ordered magnetic field. Its effectiveness is limited, however. The magnetic  field needed to drive the wind reduces the rotation of the disk, creating a gravitational barrier which the wind has to overcome  (Ogilvie \& Livio  1998, 2001, Lizano \& Galli 2015). As a result the stronger fields that would work best for accelerating a flow can not carry a significant mass or angular momentum flux.

More important is the fact that a disk supported  predominantly by rotation cannot rotate very far from Keplerian. Any angular momentum lost through a wind just causes the orbits to shrink to smaller ones which again rotate at the local Keplerian rate. Losses through a disk wind increase the mass accretion rate, but do not change the specific angular momentum of the mass accreting on the protostar. They just shorten the accretion time.

\subsubsection{stellar winds}
During most of its pre-MS life the envelope of a star is likely to be convective, hosting a dynamo-generated magnetic field as observed in MS solar type stars. A stellar wind powered by this field would likewise produce a spindown torque on the star. Since it is powered by differential rotation, the strength of the dynamo and the wind decrease as rotation decreases. At a period of 15d as seen in some young solar type stars, the spindown time scale as inferred from field stars and older clusters is much longer than the pre-MS life of stars. For stellar winds to be relevant for explaining spindown during star formation, a significant modification of the wind theory has to be devised (for an example see Matt \& Pudritz 2005).

\subsection{Loss of angular momentum and loss of magnetic flux}
\label{loss}
From the above I conclude that once they have acquired their mass, stars cannot spin down to the periods seen at the slow-rotating end of the observed distributions. This means that the explanation of their slow rotation must lie in the formation phase of the stars. For the case of the magnetic Ap stars, this has been proposed by St{\c e}pie{\'n} (2000, but see discussion in \ref{wind} above). 

The alternative explanation pursued in the following is that these stars did not have to spin down in the first place, because they were  able to accrete mass without collecting significant amounts of angular momentum. To see how this is possible, take a step back and consider the two classical `conservation' problems facing the formation of a star: the magnetic flux and the angular momentum problems. 

The magnetic flux per unit mass and the random velocities in star forming clouds are much higher than  a newly forming star can accommodate if magnetic flux and angular momentum were conserved during accretion (Mestel \& Spitzer 1956).  The accreting gas has to leave behind almost all of both these quantities in the course of the star forming process. The common view treats this (if only implicitly) as a two-step process. The magnetic flux of the gas leaves first (by ambipolar diffusion),  its angular momentum then creates an accretion disk which solves the angular momentum problem by turbulent exchange in the disk. This picture agrees  with the abundant evidence for the existence of accretion disks around young stars. But it leaves the wide range of rotation rates unexplained, and the very long period end of the range a real mystery. The picture changes significantly when one allows for some of the magnetic flux to remain in the disk.  A \textit{magnetically dominated} accretion process can then take the place of a classical accretion disk. It adds a channel for angular momentum loss that is absent in a standard accretion disk (see Mestel, 1965-I p161 item 4, and 1965-II).

\section{Magnetically dominated accretion}
\label{msa}
\begin{figure}
\center\includegraphics[width=0.4\textwidth]{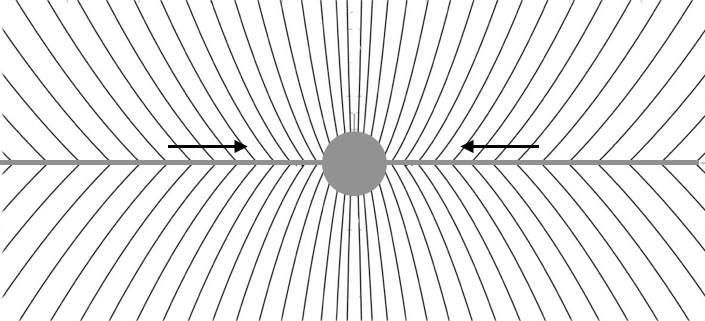}
\caption{Accretion (arrows) in a magnetic field configuration which (partially) supports the gas (grey) against the gravity of the protostar. Flow across the field is made possible by the kink in the field lines, which is subject to magnetic diffusion and/or interchange instabilities.}
\label{sketch}
\end{figure}
Take an axisymmetric configuration ($B_r,B_z$) of a field threading a disk, as sketched in Fig.~\ref{sketch}. This  field would be inherited from a star forming cloud (as opposed to generated internally in the disk).  The midplane has a surface mass density $\Sigma(r)$, which orbits at a rate $\Omega(r)$.   As in the  thin disk approximation for a standard accretion disk, neglect the gradient of gas pressure $p$, and integrate the equation of motion across the disk. For dynamical equilibrium to hold in the radial direction, this yields
\beq g(r) ={GM\,\over r^2}= g_{\rm m}+\Omega^2r,\eeq
where $M$ is the mass of the central object, and $g_{\rm m}(r)$ the (outward) magnetic acceleration due to  the bend of the field lines at the midplane:
\beq g_{\rm m}={B_rB_z\over 2\pi\Sigma}.\eeq
If this magnetic acceleration contributes significantly to support against gravity, I call the disk  \textit{magnetically dominated}, rotating at a rate less than the Keplerian rate $\Omega_{\rm K}(r)$ (Bisnovatyi-Kogan \& Ruzmaikin 1974, see also Mestel 1965-II). Such a disk is also called a `pseudodisk' (Galli \& Shu 1993).  $g_{\rm m}/g$ is then a dimensionless  measure of the degree of magnetic support. It is less than 1, since in equilibrium the  magnetic stress cannot not exceed the gravitational force.

For economy in notation write $B_rB_z=\bar B^2$. At values of the inherited field below equipartition with the gas pressure, $\bar B^2 /8\pi < p$, its contribution to support against gravity is negligible in thin disks (where $p<<\rho v_{\rm K}^2$). In this case magnetorotational (MRI) instability can plausibly generate magnetic turbulence  in the disk (provided the degree of ionisation is sufficient, see the review in Turner+ 2014). Once $\bar B^2$ is larger than $\tilde p$,  where $\tilde p$ is the gas pressure in the disk in the absence of the inherited field\footnote{this correction to $p$ is necessary because the gas pressure in the disk is modified by the pressure exerted on it by the radial component of the field,  cf.\ Ogilvie \& Livio 1998.}, such turbulence is suppressed, and the disk is in the magnetically dominated regime.  In cool disks the range of field strengths for accretion to be magnetically dominated is therefore large:
\beq \tilde p < \bar B^2/2\pi < \rho\, v_{\rm K}^2.\eeq
Numerical simulations of MHD jets from accretion disks nicely show how both forms of accretion flow can coexist  (e.g.\ McKinney \& Gammie 2004, Liska+ 2018).  A turbulent magnetorotational disk behaving like a standard accretion disk can surround a region of magnetically dominated accretion (the `flux bundle' producing the jet).  Magnetorotational turbulence is absent in the magnetically dominated region; instead, accretion there  takes the form of `fingers' penetrating between field lines. The term `magnetically arrested' accretion  is also used for this region (Igumenshchev+ 2003). The two regions are quite distinct in the simulations. 

\subsection{Mass transport}
\label{transport}
In the traditional view, mass accretes across field lines through ambipolar diffusion (Mestel \& Spitzer 1956). Since this process is governed by collisions between ions and neutrals it scales with the inverse square of the mass density, for a given degree of ionisation,  and is therefore most effective at the low gas densities of molecular cloud cores.  If the magnetic flux remaining trapped in the accreting gas is sufficient to establish a magnetically dominated pseudo-disk, an interchange instability can become competitive with microscopic diffusion processes, especially in denser regions closer to the protostar. 

In a configuration as in Fig.\ \ref{sketch}, magnetic tension due to  the kink at the midplane $z=0$ supports mass against gravity in the form of a sheet  with surface mass density $\Sigma$. Such a configuration can be unstable, with the weight of the magnetically supported gas driving exchange between magnetic field and surface density. The  presence of the process depends only on the magnetic field and gravity; at low plasma-$\beta$ it is independent of the  gas pressure. The instability develops small length scales in the $\phi$-direction (perpendicular to the plane of Fig.~\ref{sketch}) (Spruit \& Taam 1990, Spruit+ 1995). 
It is closely analogous to two-dimensional Rayleigh-Taylor instability in ordinary hydrodynamics (e.g.\ the numerical simulations in Stehle+ 2001). Since surface density per unit of magnetic flux is conserved  (in ideal MHD), the condition for instability is that the ratio $\Sigma/B_z$ increases with distance (opposite to gravity):
\beq \partial_r(\Sigma/B_z)>0. \label{RTic}\eeq 

Rayleigh-Taylor interchange in a magnetically dominated disk takes place by field lines exchanging positions in the $r-\phi$ plane.  Overturning flows in the $r-z$ plane, which would bend field lines, are suppressed by magnetic tension\footnote{In a more isotropically turbulent disk, as in MRI turbulence, this is different. In that case rapid diffusion of the radial field component across the disk thickness leads to the well-known difficulty of advecting magnetic flux inward through a standard thin accretion disk (c.f.\ van Ballegooijen 1989, Bardou \& Heyvaerts 1996.) This problem does not occur with RT interchanges, because they work on the radial gradient scale instead of the vertical (disk thickness) scale.}.

A nearby example of magnetic RT exchange that has been studied in great detail is the so-called quiescent prominence in the solar atmosphere (e.g.\ Anzer 1969, Su+ 2015). A sheet of cold ($\sim 10^4$ K) gas, fed by condensation from the surrounding hot ($2\,10^6$ K) corona, is suspended against gravity by a kink in the field lines as in Fig.~\ref{sketch}.  When such a prominence sheet is seen face-on, the interchange process can be observed in detail\footnote{for a nice example see this video by Paolo Porcellana, \url{https://www.youtube.com/watch?v=X6sUSiqLYLg}}, showing dense (brighter) gas flowing down, exchanging with upward moving less dense gas (dark, more strongly magnetic). The horizontal length scales of the process are much smaller than the vertical, which extends almost over the entire visible height of the prominence. The vertical velocities are some 10\% of the free fall speed over the height of the prominence. 

\subsection{Angular momentum loss}
\label{Jloss}

Through the magnetic flux it inherited, the accreting magnetically dominated pseudodisk is still connected to its birth cloud. As long as it has not lost all its magnetic flux this gives the disk a channel for angular momentum loss that is absent for a star interacting only with an accretion disk (c.f.\ Mestel 1965-I). To make this a plausible possibility, consider a field configuration as sketched in Fig.\ \ref{sketch}. 

Near the origin the field is strong enough to corotate with the rate $\Omega(r)$ of the midplane, but at some (possibly large) distance it starts feeling the connection to the birth cloud. The difference in rotation rate with the slowly rotating cloud core causes the field to develop an azimuthal component $B_\phi$ (`winding up').  The strength of the field  decreases with distance from the midplane, and at some point the field lines are unable to force the gas to continue rotating with their footpoint rate $\Omega$. This happens when the rotation velocity $\Omega\varpi$ approaches the Alfv\'en speed $B/\sqrt{4\pi \rho_{\rm c}}$, where $\rho_{\rm c}$ is the mass density in the cloud core, and $\varpi$ the distance from the rotation axis.  At the location where this happens, the Alfv\'en surface, the azimuthal field $B_\phi$ becomes comparable with the poloidal component $B_{\rm p}=(B^2_\varpi+B_z^2)^{1/2}$. At this distance the potential field approximation does not apply any more. Beyond it the field behaves instead like a standing Alfv\'en wave, transmitting rotational energy from the disk into the cloud.  

Up to the Alfv\'en surface, the poloidal field component is close to a potential configuration (as in the analogous case of a magnetic wind (c.f.\ Mestel 2012 Ch.\ 7.2 or the summary in Spruit 1996). Azimuthal equilibrium  (torque balance)  shows that the quantity $\varpi B_\phi$, the `torque per unit magnetic flux' is constant along a field line.
If the  Alfv\'en distance is large, the torque is small. Near the midplane the angle $B_\phi/B_{\rm p}$ is then small, and a potential field a good approximation. This  simplifies estimates of angular momentum loss, since useful potential field configurations are available in the literature. Details are given in the next sections.

\subsubsection{The last bit of rotation}
\label{lastbits}
The angular momentum of slowly rotating stars (periods of weeks, say) is small.  Even a weak magnetic connection between the accreting matter and the distant Alfv\'en surface in the cloud core may suffice to remove the last bit of angular momentum from the gas before it accretes onto the protostar. A magnetic connection can exist only during the accretion process, however, since the strength of the connecting field is maintained by the balance between advection and outward diffusion of field lines in the pseudodisk feeding the star. Once accretion stops, diffusion dominates and the magnetic field slips out of  the disk. From that point on the angular momentum of the star is frozen unless accretion were to set in again. Until this happens the star has only its own field connecting to the outside (whether dynamo-generated or accreted in the course of its formation). As discussed in sect.~\ref{brake} magnetic braking is then not enough to explain the observed very long periods. Instead, very slowly rotating stars must have remained connected to their birth cloud throughout their formation, including the last phases of mass accretion. This may be a rare occurrence. Then again, stars with periods of decades are also rare: a few percent of the Ap stars, which themselves only make up some 10\% of the A star population.

\subsection{Angular momentum-free accretion}

As argued above, rotation periods of decades and longer as observed in Ap stars must result from the accretion of mass without angular momentum. What are the conditions for this to be possible? In a Lagrangian frame following a mass element drifting through the disk the element's angular momentum $J$ is a function of time. The torque acting on the mass reduces $J$ in a way that depends on rotation rate, but also indirectly on factors like the distance to the Alfv\'en surface, which varies with the distance of the mass element from the accreter. Suppose in some range of distances from the accreter we can approximate the effect of the torque (still to be calculated in detail in the following) with a simple fitting formula,
\beq
{{\rm d}J\over {\rm d}t}=-k\,J^\alpha,\label{dom}
\eeq
with parameters $k$ and $\alpha$. Integration yields 
\beq
J/J_0=[1-(1-\alpha)\,t/t_{\rm c}]^{1/(1-\alpha)},
\eeq
where $J_0$ is the initial value of $J$, and $t_{\rm c}=J_0^{1- \alpha}/k$. If $\alpha$ is larger than $1$, $J$ declines smoothly for $t\rightarrow\infty$. If $\alpha<1$, however, $J$ vanishes at a finite time $t_0=t_{\rm c}/(1-\alpha)$, corresponding to a finite distance inside the outer boundary (where the mass is at $t=0$). This shows that angular momentum can be extracted completely if $\alpha<1$. In other words, this happens if the torque does not decrease too quickly with decreasing rotation rate.  To see if  this property of the torque is also realized in a model with more physics requires some calculations.

\section{Accretion model}
\label{magacc}

To make the model tractable assume a steady state, such that the inward advection of magnetic field by the accretion flow is balanced by outward diffusion of the field lines. 
The accretion of mass from the cloud is represented by a constant mass flux $\dot M$ entering at the outer boundary of the model. As discussed in (\ref{wind}), the field above the midplane is a potential field determined by the distribution $B_z(r)$ of its vertical component on the midplane. Angular momentum loss from the disk is calculated from the torque acting at the Alfv\'en surface in the birth cloud, as described in sect.~\ref{Jloss}. The cloud is assumed to have a  uniform mass density $\rho_{\rm c}$.  The (axisymmetric) cylindrical coordinates are ($\varpi, z $) while $r$ will be used to denote  radial distance for quantities defined on the midplane. The angular momentum balance, evaluated at the midplane, is
\beq 
\partial_r(2\pi r\Sigma v_r\,\Omega r^2)+ 2\pi r T=0,\label{angbal0}
\eeq
where $v_r$ is the accretion velocity (taken positive), and  $\Sigma$ the surface mass density. The torque $T$ per unit surface area (not to be confused with the total torque $\cal T$ in sect.\ \ref{brake}) follows from the $r\phi$ component of  the magnetic stress:
\beq T(r,\Omega)=r{B_{\rm p} B_\phi\over 2\pi}, \label{torq0}\eeq
 where $B_{\rm p}(r)$ is the poloidal field amplitude $(B_r^2+B_z^2)^{1/2}$, and the factor $2\pi$ comes about because the torque acts on both sides of the disk. Since the accretion rate $r\Sigma v_r$ is assumed constant, (\ref{angbal0}) yields
\beq
\partial_r(\Omega r^2)+{T\over \Sigma v_r}=0. \label{angbal1}
\eeq
Since $v_r$ will be assumed small compared with the azimuthal velocity, and gas pressure is neglected, radial balance of forces is between gravity, the centrifugal force and the radial magnetic stress is:
\beq 
\Sigma(g-\Omega^2 r)={B_rB_z\over 2\pi},\label{req}
\eeq
where $g(r)$ is the acceleration of gravity by the central mass. Using this, the torque (\ref{torq0}) can be written as 
\beq 
T(r,\Omega)=\Sigma r(g-\Omega^2 r)\,q{B_\phi\over B_{\rm p}},
\eeq
where q=$B_r/B_z+B_z/B_r\approx 2.13$ (see Appendix),
and the angular momentum balance at the midplane (\ref{angbal1}) becomes
\beq 
\partial_r(\Omega r^2)+{r\over v_r}(g-\Omega^2 r)\,q{B_\phi\over B_{\rm p}}=0. \label{angbal2}
\eeq
The angle $B_\phi/B_{\rm p}$ that determines the torque follows from the connection with the cloud. In a force free field the `torque per field line' is constant along a field line. In an axisymmetric field, this is equivalent to $ \varpi B_\phi$ being constant along a field line. If $\varpi_{\rm A}$ is the distance of the field line from the axis at the Alfv\'en distance (sect. \ref{brake}), this means
\beq
r B_\phi=\varpi_{\rm A}B_{\phi,{\rm A}}.
\eeq
As in the case of a magnetic stellar wind, $B_{\phi,{\rm A}}\approx - B_{\rm p,A}$ at the Alfv\'en distance, so at the disk surface
\beq B_{\phi}/B_{\rm p}\approx-{\varpi_{\rm A}\over r} {B_{\rm p,A}\over B_{\rm p}}. \label{azangl}\eeq
This is a small angle, justifying a potential field approximation above the disk, provided $B_{\rm p}$ decreases more rapidly than $\varpi^{-1}$, measured along a field line. The Alfv\'en distance is defined as the  location where the rotation of the field line equals the Alfv\'en speed,
\beq  \Omega \varpi_{\rm A}={B_{\rm p,A}\over (4\pi\rho_{\rm c})^{1/2}},\label{corot}\eeq
where $\rho_{\rm c}$ is the mass density of the birth cloud. For simplicity and to avoid adding additional model parameters, I assume it to be constant throughout. For a given rotation rate and a known field configuration $B_{\rm p}(\varpi,z)$, (\ref{corot}) can be solved for $\varpi_{\rm A}$ as a function of the footpoint distance $r$ of the field line, yielding the azimuthal field angle at the midplane (\ref{azangl}) and the torque term in (\ref{angbal2}).

\subsection{Scale-free accretion}
\label{scalefree}

Accretion from a cloud core at sub-parsec scales to a protostar covers 6-7  orders of magnitude in length scale. For most of these orders, information about strengths and configuration of the magnetic field still appears to be limited. In view of this I adopt a scale-free model that treats the orders of magnitude in a more generic way. In such a model the physical quantities scale as powers of distance $r$ from the star. The gravitational acceleration scales as $1/r^2$, and velocities as $r^{-1/2}$, which I also assume to be the case for the accretion velocity $v_r$ in eq. (\ref{angbal2}):
\beq v_r=\epsilon (r\,g)^{1/2},\label{vr}\eeq
where the fraction $\epsilon$ is an adjustable parameter of the model. With the accretion rate $\sim r\Sigma v_r={cst}$, the surface density then scales as
\beq \Sigma\sim r^{-1/2}. \eeq
From radial equilibrium (\ref{req}),
\beq B_rB_z\sim r^{-5/2}, \eeq
at least in those regions where magnetic support against gravity dominates over the centrifugal acceleration. For the configuration to be scale-free, $B_r$ and $B_z$ have to have the same dependence, $B_{r,z}\sim r^{-5/4}$. 
The potential field ${\mathbf B}(\varpi,z)$ for this  distribution of $B_z$ on the midplane can be written in terms of a hypergeometric function of argument $\varpi^2/z^2$ (Sakurai 1978, reproduced here in the Appendix). In agreement with its scale-free nature, it is sufficient to compute a single field line, all others differ only in field strength and length scale. The field line angle $B_r/B_z$ on the midplane is a constant. 

These scalings are the same as in a scale-free hydrodynamic accretion disk. Since RT instability is driven by gravity it naturally has the scaling (\ref{vr}). The mass-to-flux ratio $\Sigma/B_z$ then scales as $r^{3/4}$, thereby satisfying the condition for RT instability (\ref{RTic}). These properties make RT useful in a scale free accretion model, especially in regions where it would operate faster than diffusive processes.

Since the model is scale-free, there is no place in it for the accreting protostar.  Instead, it has a singularity at the origin. Its effect is limited, however, since the mass $ r^2\Sigma\sim r^{3/2}$ and magnetic flux $ r^2 B_z\sim r^{3/4}$, are regular at the origin. The field strength is largest at small distance $r$, but most of the magnetic flux is at large distances. This combination gives the field configuration the `hourglass' or collimating  shape expected from a field that is being concentrated by a disk-like accretion flow. 

It helps to write the angular momentum equation in dimensionless form. Write
\beq x=r/r_0,\quad y(x)=\Omega/\Omega_{\rm K},\quad \alpha_{\rm t}=-B_\phi/B_{\rm p},\eeq
where $ r_0$ is the starting point for the integration of eq.\ (\ref{angbal2}), and $\Omega_{\rm K}$ the Kepler rotation rate $\sim x^{-3/2}$. 
The `torque factor' $\alpha_{\rm t}(x,y)$ (positive) measures the (absolute value of) the braking torque.\ It is determined by the solution of the corotation equation (\ref{corot}). With the dimensionless accretion velocity $\epsilon$ (eq.\ \ref{vr}) the angular momentum equation (\ref{angbal2}) can be written as
\beq x\partial_x(y)+ y/2- (1-y^2)\,q\,\alpha_{\rm t}(x,y)/\epsilon=0.\label{angbalf}\eeq
The model has three parameters: the initial value $y_0$ at $x=1$, the value of the  dimensionless accretion velocity $\epsilon$, and a parameter for the conditions at the corotation point in the cloud. For this last parameter take the height $z_{\rm c}$ along the field line, measured at the Alfv\'en distance, in units of $r_0$. It is a function of the rotation rate of the field line, the mass density at the corotation point, and the field strength there. The rotation of the cloud itself is assumed to be the (very slow) Keplerian rate at the corotation point.

\subsection{Computations}

\begin{figure}
~~~\includegraphics[width=0.3\textwidth]{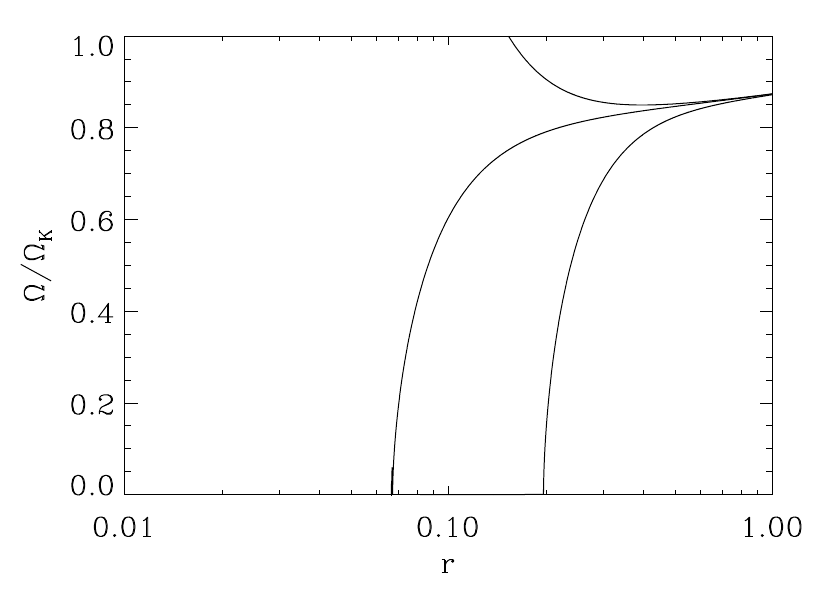}\\
\includegraphics[width=0.32\textwidth]{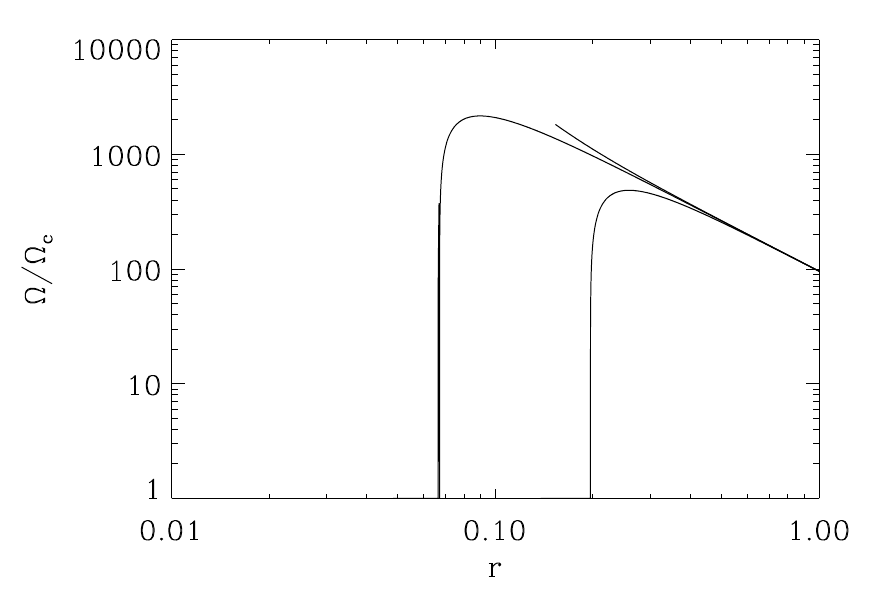}\\
\includegraphics[width=0.32\textwidth]{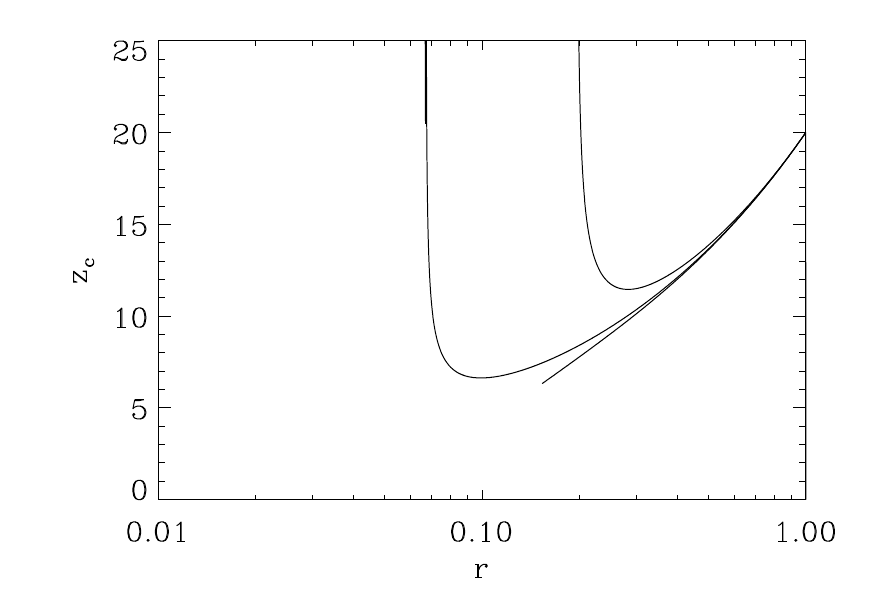}\\
\caption{\label{sol}Top:  rotation $\Omega$ as a function of distance $r$ from the origin, in units of the Keplerian rate $\Omega_{\rm K}$, for initial values 0.8660, 0.8656, and 0.8640 (top to bottom). Accretion velocity parameter $\epsilon=0.1$. Middle panel: corresponding rotation rates in units of the cloud value $\Omega_{\rm c}$. Bottom panel: corotation height $z_{\rm c}$ in the cloud. }
\end{figure}

Eq.\ (\ref{angbalf}) is integrated inward from $x=1$, $y=y_0$ using a semi-implicit Euler scheme. Fig.\ \ref{sol} shows an example for the case $\epsilon=0.1$ and $z_0=20$. The outcome depends sensitively on the assumed initial rotation rate $y_0$. Above a critical value around 0.85, the increase of the rotation rate by angular momentum advection (first 2 terms in \ref{angbalf}) wins over the braking torque. Rotation then reaches the Keplerian value $y=1$ within a finite distance, at which the calculation is terminated. Already at a slightly lower  initial rotation rate, the braking torque wins, and the rotation quickly drops to values near the rotation rate of the cloud. This `bifurcation' behavior is as suggested by the simple example in sect.\ \ref{lastbits}. 

As an example of what the results mean for actual star formation conditions, assume accretion  from a distance $r_0$ of 20 AU of mass initially rotating at 80 \% of the Keplerian rate, at a rate $\dot M=10^{-5} M_\odot/{\rm yr}$ on a $1\, M_\odot$ protostar. The surface density at $r_0$ is then $\Sigma\approx 50$ g/cm$^2$. Magnetic support of the remaining 20\% of gravity requires $B_rB_z/2\pi=0.2 g$, which yields a field strength $\vert B\vert\approx 150$ mG at $r_0$. The field strength at the surface of the protostar of radius $10^{11}$ cm would then be about 7 kG. 

\subsubsection{consistency}

The formulation above starts with the assumption of a scale-free dependence on distance.  The angular momentum equation (\ref{angbal2}) is consistent with this scaling only as long as the centrifugal acceleration in the radial balance can be treated as a correction, so that the rotation decouples from the radial balance. This is justified in the slowly rotating parts of the solutions that are of most interest, but fails near the starting point of the integration, where the centrifugal acceleration makes up 85\% of support against gravity, in the example given above. 
The results should therefore not be taken quantitatively in this region. The main result of the model: the `bifurcating' nature of the solutions including the possibility of complete spindown, should be unaffected by this uncertainty.

\section{Discussion}
After eliminating other possibilities, I find that the existence of extremely slowly rotating stars suggests a mode of star formation whereby mass can accrete without angular momentum. The essential ingredient is that some magnetic flux from the birth cloud survives in the accretion process, such that a sustained magnetic connection exists between the accretion flow and the birth cloud. The model presented here shows that this is possible even when the connecting cloud is at a very large distance from the accreter. The physics involved has in essence been developed already in the 1960's\footnote{Mestel (1965-I p161) writes as a starting assumption: "As long as the field lines retain a suitable structure, the magnetic field can transport angular momentum from a central condensation to the surrounding gas".}  

The calculations presented show that the boundary in parameter space between the two possible outcomes for the rotation of the accreting gas (Keplerian or the rotation rate of the cloud) is quite sharp (a `bifurcation'). This does not explain the range of intermediate rotation rates in young solar type stars and Ap stars. The star formation process is likely to be episodic, however. Intermediate rotation rates would result if the balance tips a few times between episodes. The very slowest rotation rates would then occur if  cloud rotation is the outcome of all episodes.  The observed frequency of the slowest stars then gives an indication how many episodes are involved (not more than a few, probably).

The advent of powerful numerical simulations and critically new observations has revived discussions of the relation between angular momentum and magnetic flux in star formation. Simulations often find significantly sub-Keplerian accretion modes (Allen+ 2003, Hennebelle \& Ciardi 2009, K{\"u}ffmeier+ 2017, Chen \& Ostriker 2018, Wuster+ 2018). The  phenomenon is sometimes named the `magnetic braking catastrophe'. Formation of the expected Keplerian accretion disk seems to require additional ingredients, such as inclusion of a Hall effect in the simulations (Kwon+: 2018) or inclusion of turbulence (Seifried+ 2015). 

At the same time,  Keplerian rotation does not seem to be observed as ubiquitously as expected. Lee+ 2018, for example, report an object showing  approximately Keplerian motions from 60 to 130 AU without any indication of Keplerian orbits within 60 AU.

The results presented here show that accretion with low angular momentum can be a natural part of the star formation process rather than a catastrophe to be avoided.

\begin{appendix}
\section{ Field configuration}

\noindent
In cylindrical coordinates ($\varpi, \varphi, z$) an axisymmetric potential field with vertical comonent $B_z$ varying as $\varpi^{-\mu}$ on the midplane can be written as (Sakurai 1987)
\bea
B_\varpi=&{\varpi\Gamma(\mu+1)\over 2 z^{\mu+1}}F({\mu+1\over 2},{\mu+2\over 2},2,-{\varpi^2\over z^2}),\quad z>0, \\
=&{2^{\mu-1}\over\varpi^\mu}\Gamma({\mu+1\over 2})/\Gamma(1-{\mu-1\over 2}),\quad z=0,\\
B_z=&{\Gamma(\mu)\over z^\mu}F({\mu\over 2},{\mu+1\over 2},1,-{\varpi^2\over z^2}),\quad z>0,\\
= & {2^{\mu-1}\over\varpi^\mu}\Gamma({\mu\over 2})/\Gamma(1-{\mu\over 2}), \quad z=0.
\eea
The hypergeometric function $F$  and  gamma function $\Gamma$  are efficiently evaluated with the routines given in Numerical Recipes. Since the configuration is scale invariant, only a single field line needs to be computed explicitly. The angle $B_\varpi/B_z$ on the midplane $z=0$ is a constant depending only on $\mu$. For $\mu=1$ it is equal to 1, for $\mu={5/4}$ it is $\Gamma(9/8)/\Gamma(7/8)\cdot\Gamma(3/8)/\Gamma(5/8)\approx 1.42813$.  For $\mu<2$ the magnetic flux is dominated by the outer regions, giving the configuration an hourglass or `collimating' shape.

\end{appendix}

\end{document}